\documentclass[12pt]{iopart}
\expandafter\let\csname equation*\endcsname\relax
 \expandafter\let\csname endequation*\endcsname\relax
\usepackage{amsmath,amssymb,eucal,graphicx,graphics,float}
\begin{document}
\title{Distinct Degrees and Their Distribution in Complex Networks}

\author{P. L. Krapivsky}
\address{Department of Physics, Boston University, Boston, MA 02215, USA}
\author{S. Redner}
\address{Center for Polymer Studies and Department of Physics, Boston University, Boston, MA 02215, USA}

\begin{abstract}
  We investigate a variety of statistical properties associated with the
  number of distinct degrees that exist in a typical network for various
  classes of networks.  For a single realization of a network with $N$ nodes
  that is drawn from an ensemble in which the number of nodes of degree $k$
  has an algebraic tail, $N_k\sim N/k^\nu$ for $k\gg 1$, the number of
  distinct degrees grows as $N^{1/\nu}$.  Such an algebraic growth is also
  observed in scientific citation data.  We also determine the $N$ dependence
  of statistical quantities associated with the sparse, large-$k$ range of
  the degree distribution, such as the location of the first hole (where
  $N_k=0$), the last doublet (two consecutive occupied degrees), triplet,
  dimer ($N_k=2$), trimer, etc.
\end{abstract}

\section{Introduction}

A complete microscopic representation of a macroscopic system is usually
unavailable and often unnecessary, especially if the system is evolving or it
is taken from an ensemble and the goal is to understand the typical features
of the ensemble.  Thus instead of determining a huge number of parameters
(such as the $10^{23}$ coordinates and momenta of atoms), it often suffices
to know a few useful macroscopic quantities (like the total number of atoms
and the total energy) to understand the bulk properties of a macroscopic
system.  

In the realm of networks, one usually starts with an ensemble of large
networks that are generated according to a specified and not completely
deterministic algorithm.  In analogy with other bulk systems, we are
typically interested in macroscopic-like network characteristics, such as the
total number of links, the total number of triangles, the total number of
clusters (maximal connected components), etc.~\cite{Newman}.  Two of the most
useful macroscopic characteristics are the cluster-size distribution and the
degree distribution.

The degree of a node (the number links attached to the node) is perhaps the
simplest local network characteristic.  It has been now been extensively
studied, with an emphasize on networks with broadly distributed
degrees~\cite{BA02}.  Here we analyze the number of \emph{distinct} degrees
$D_N$ that exists for a given network of size $N$. The number $D_N$ varies
from realization to realization, but for the ensembles that we study $D_N$
turns out to be a self-averaging quantity, so that its mean value is the most
important characteristic.  We focus on $\langle D_N\rangle $ which we
generally write as $D_N$ when no ambiguity is possible.

\begin{figure}[ht]
\begin{center}
\includegraphics[width=0.325\textwidth]{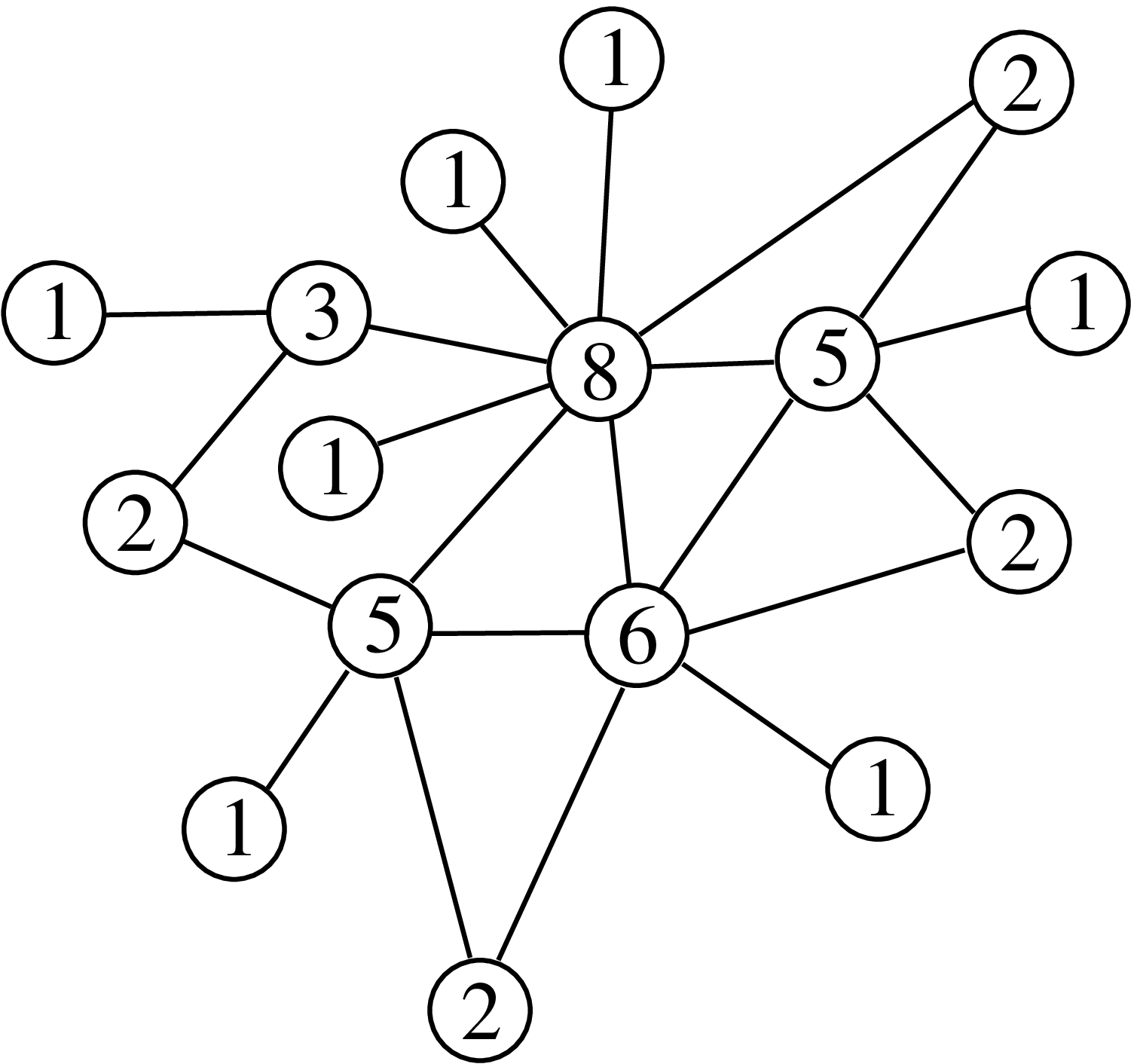} 
\caption{A network of 16 nodes, with node degrees indicated.  In this
  example, $N_k=\{7,4,1,0,2,1,0,1\}$.  The number of distinct degrees
  $D_{16}=6$, the last doublet occurs at $k=5$, the last dimer also at $k=5$,
  the first hole at $k=4$, and $k_{\rm max}=8$.}
  \label{sketch-small}
\end{center}
\end{figure}

We also investigate the locations of the first hole (the smallest $k$ where
$N_k$ equals zero), the last doublet (the largest $k$ value for which $N_k>0$
and $N_{k+1}>0$), last triplet, the last dimer (the largest $k$ value where
$N_k=2$), trimer, etc.\ in the degree distribution (Fig.~\ref{sketch-small}).

\begin{figure}[ht]
\begin{center}
\includegraphics[width=0.5\textwidth]{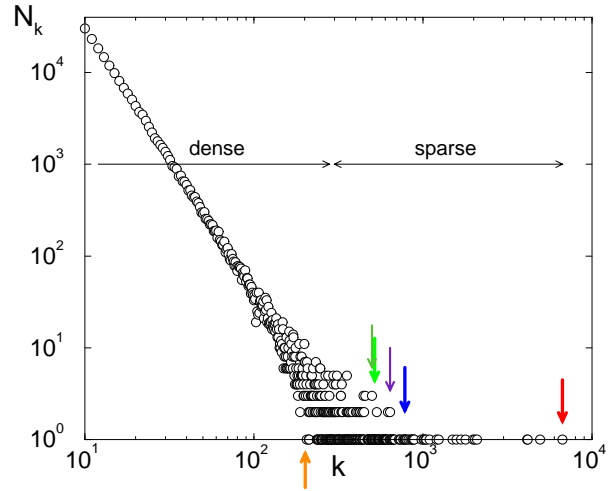} 
\caption{Number of nodes of degree $k\geq 10$ for a single network
  realization of $N=10^7$ nodes that is grown by strictly linear preferential
  attachment.  The largest degree is $k_{\rm max}=6693$, $D_N=465$, the last
  doublet occurs at $k=782$, the last dimer at $k=641$, the last triplet at
  $k=518$, the last trimer at $k=500$, and the first hole at $k=201$
  (arrows).}
  \label{sketch-10000000}
\end{center}
\end{figure}

The number of distinct degrees $D_N$ exhibits interesting behavior for
network ensembles in which the degree distribution has an algebraic tail;
hence we focus on such networks.
%The precise definition of an ensemble plays little role.  
For concreteness, we consider networks that are grown by preferential
attachment.  The best-known case is strictly linear preferential
attachment~\cite{Y25,S55,P65,BA99,KRL00,DMS00}, in which a new node attaches
to a pre-existing node of degree $k$ with rate $A_k=k$.  To illustrate the
quantities studied here, we plot the degree distribution for a realization of
such a network of $N=10^7$ nodes (Fig.~\ref{sketch-10000000}).  For small
$k$, every degree is represented, that is, $N_k>0$.  As $k$ increases,
eventually a point is reached where $N_k$ first equals zero; this defines the
first ``hole'' in the degree distribution.  Holes become progressively more
common for larger $k$ and eventually the distribution becomes sparse.
Figure~\ref{sketch-10000000} also indicates the position of the last doublet,
the largest $k$ for which $N_k>0$ for two consecutive $k$ values, while the
last dimer is defined as the largest $k$ value for which $N_k=2$.  One can
analagously define the last triplet and last trimer, etc.  As $k$ continues
to increase, the degree distribution is non-zero at progressively more
isolated $k$ values and eventually the distribution terminates when largest
network degree $k_{\rm max}$ is reached.

One of our principal results is that
\begin{equation}
\label{D_exact}
D_N \simeq  \Gamma\big(1-\tfrac{1}{\nu}\big)\,(RN)^{1/\nu}
\end{equation}
for networks whose degree distribution has the algebraic tail
\begin{equation}
\label{Nk}
N_k\simeq NR k^{-\nu}\qquad\text{when}\quad k\gg 1,
\end{equation}
where $R$ is a constant of the order of 1.  

The behavior of $D_N$ parallels that of Heap's law of
linguistics~\cite{H60,H78}, in which the number of distinct words in a large
corpus of $N$ words grows sub-linearly with $N$.  Recent
work~\cite{LW05,LZZ10,GA13} has related the $N$ dependence in Heap's law to
the dependence of word frequency versus rank in this same corpus --- Zipf's
law~\cite{Z36}.  Because of the simplicity and explicitness of scale-free
network models, we can quantify the statistical properties of $D_N$ more
precisely than in word-frequency statistics.  It is also worth noting that
the number of distinct degrees in a particular realization of a network is
reminiscent of the ``graphicality'' of a network.  Namely, given a set of
disconnected nodes, each with a specified degree, one can ask which degree
sequences allow all the nodes to be connected into a single component without
multiple links between the same nodes~\cite{EG60,DKTB10,DGB11}.  The number
of distinct degrees provides complementary information oabout which degree
sequences are actually realized in a complex network.

\section{Distinct Degrees}

Consider networks whose degree distribution has the asymptotic power-law form
of Eq.~\eqref{Nk}.  We deal only with sparse networks, for which $\nu>2$.  A
network with such a degree distribution can be easily constructed by the
redirection algorithm~\cite{KR01}, in which a new node either attaches to a
random-selected ``target'' node with probability $1-r$ or to the ancestor of
the target with probability $r$.  This algorithm generates a scale-free
network whose growth rule is precisely shifted linear preferential
attachment, with the attachment rate to a node of degree $k$,
$A_k=k+\lambda$, and with $\lambda=\frac{1}{r}-2$.  This growth rule leads to
a degree distribution that has the form \eqref{Nk} with exponent
$\nu=1+\frac{1}{r}$.  We use this redirection algorithm for our simulations
and interchangeably refer to the growth mechanism as either shifted linear
preferential attachment or redirection.

\begin{figure}[ht]
\begin{center}
\includegraphics[width=0.5\textwidth]{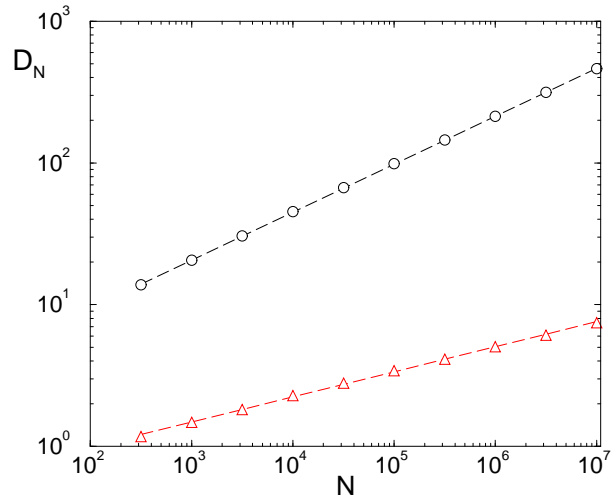}
\caption{The average number of distinct degrees $D_N$ versus $N$ for networks
  that are grown by redirection with redirection probability $r$.  The upper
  curve ($\circ$) corresponds to $A_k=k-\frac{1}{2}$ or to redirection
  probability $r=\frac{2}{3}$.  Here the degree distribution exponent is 5/2
  and $D_N=BN^{2/5}$, with
  $B=(3/2)^{2/5}\pi^{-1/5}\Gamma(3/5)=1.393019\ldots$.  The lower curve
  ($\bigtriangleup$) corresponds to $A_k=k$ or $r=\frac{1}{2}$.  Here $D_N$
  is given by \eqref{B_exact}.  Each data point represents an average over
  $10^4$ realizations.  The dashed lines correspond to the theoretical
  prediction \eqref{D_exact}.}
  \label{DD}
\end{center}
\end{figure}

To determine the number of distinct degrees that appear in a typical
realization of a large network, first notice that for $k$ in the range $k\leq
K=(NR)^{1/\nu}$, $N_k\geq 1$.  In this dense regime of the degree
distribution (Fig.~\ref{sketch-10000000}), all degrees with $k<K$ are
present.  This range therefore gives a contribution of $(NR)^{1/\nu}$ to
$D_N$.  In the complementary sparse range of $k>K$, we estimate the number of
distinct degrees, by integrating the degree distribution for $k>K$.  Adding
the contributions from the dense and sparse regimes gives
\begin{equation}
\label{D_naive}
D_N^{\text{naive}} = \frac{\nu}{\nu-1}\,K, \qquad\qquad K = (NR)^{1/\nu} \,.
\end{equation}
While the $N$-dependence is correct, $D_N\sim N^{1/\nu}$, the amplitude is
wrong.  A better estimate can be obtained by assuming that the probability
distribution for the number of nodes of each degree $k$ is the Poisson
distribution with average value $N_k$ given by \eqref{Nk}.  Then $P_k
\equiv\,\mathrm{Prob[(\#\ nodes\ of\ degree\ } k) \geq 1] = 1-\exp(-N_k)$.
Using this property leads to a more accurate estimate (this same approach was
developed in Ref.~\cite{GA13})
\begin{equation}
\label{Poisson}
D_N  = \sum_{k\geq 1} \left[1-e^{-N_k}\right] \,.
\end{equation}
Replacing the sum by an integral, we ultimately obtain \eqref{D_exact}.  For
strictly linear preferential attachment, $R=4$ and $\nu=3$, so that
\begin{equation}
\label{B_exact}
D_N=BN^{1/3}\qquad B=2^{2/3}\Gamma\left(\tfrac{2}{3}\right) = 2.149528\ldots
\end{equation}
In contrast, the naive estimate \eqref{D_naive} for the amplitude is
$B_{\text{naive}} = 3\cdot 2^{-1/3} = 2.381101\ldots$, which exceeds the more
accurate value by $\approx 11\%$.  

Generally $D_N/D_N^{\text{naive}}=\Gamma\left(2-\tfrac{1}{\nu}\right)$, so
for the admissible range of $2<\nu<\infty$, this ratio monotonically
increases from $\tfrac{1}{2}\sqrt{\pi}\approx 0.886227$ to 1.  As shown in
Fig.~\ref{DD}, simulation results are in excellent agreement with our
theoretical predictions.  A more detailed asymptotic analysis (employing
methods developed in Ref.~\cite{KR02}) indicates that the average number of
distinct degrees admits the expansion, $D_N=BN^{1/3} + C + \ldots$ for
strictly linear preferential attachment.  This allows us to extract a precise
estimate of $B$ from the data that is in excellent agreement with
Eq.~\eqref{B_exact}.

\begin{figure}[ht]
\begin{center}
\includegraphics[width=0.5\textwidth]{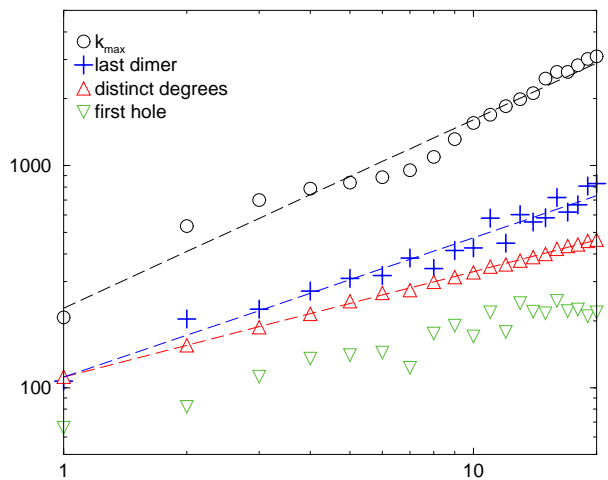}
\caption{The maximum degree (number of citations) and the number of distinct
  degrees for the Physical Review citation network during the period
  1893--2003.  Also shown are the locations of the last dimer and the first
  hole.  The dashed lines are the power-law fits with respective exponents
  0.849. 0.627, 0.474,and 0.430.  The data are measured at 20 equally-spaced
  network sizes as discussed in the text. }
  \label{max-distinct}
\end{center}
\end{figure}

The general behavior outlined above for the number of distinct degrees and
related quantities is also observed in the citation network of the Physical
Review.  Because this journal has grown roughly exponentially with
time~\cite{R05,KR05}, it is not appropriate to use publication date as a
proxy for the network size.  Since the citation data is presented as a list
of links, each in the form of \emph{citing paper} $\to$ \emph{cited paper},
it is more natural to use the chronologically-ordered number of links as the
proxy for network size.  We use the Physical Review citation data as of 2003,
which contains $L=3,110,866$ total links (citations).  The maximum network
degree (the highest-cited paper), the location of the last dimer, the number
of distinct degrees, and the location of the first hole dimer are measured
when the network size is $\frac{m}{20}L$, with $m=1,2,\ldots,20$
(Fig.~\ref{max-distinct}).

Naive power-law fits to the first three datasets in Fig.~\ref{max-distinct}
give $k_{\rm max}\sim L^{0.849}$, $D_L\sim L^{0.627}$, and $\langle
h_1\rangle\sim L^{0.474}$.  Let us provisionally assume that the citation
distribution has a power-law dependence on $L$ and, by implication, the same
dependence on $N$\footnote{While a power-law gives a reasonable visual fit to
  the data, later and larger-scale analyses~\cite{R05,WLG09,P09,PPD10,GS12}
  suggest that the citation distribution has a log-normal or stretched
  exponential behavior, rather than a power-law form.}.  Using $k_{\rm
  max}\sim N^{1/(\nu-1)}$ and the dependences for the number of distinct
degrees and location of the last dimer given in Eqs.~\eqref{D_naive} and
\eqref{V}, we infer the respective exponents for the degree-distribution
exponent values of 2.18, 2.09, and 2.11.  Thus these three properties are
internally consistent under the assumption the citation distribution has a
power-law form with exponent in the range $2.1$--$2.2$.

\begin{figure}[ht]
\begin{center}
\includegraphics[width=0.5\textwidth]{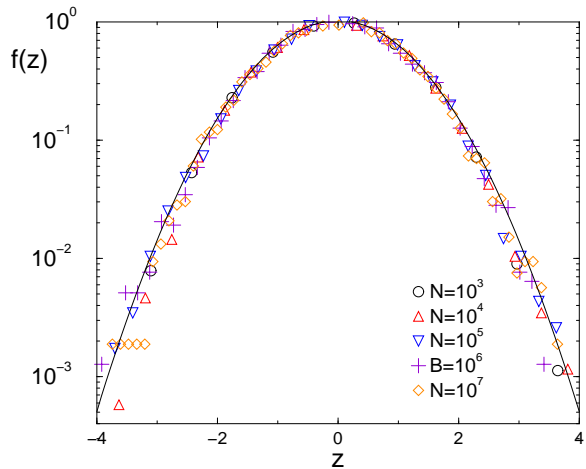}
\caption{Scaled distribution of distinct degrees $f(z)$ for $N$ up to $10^7$,
  with $10^4$ network realizations for each $N$, for redirection probability
  $\frac{1}{2}$ (corresponding to strictly linear preferential attachment).
  The smooth curve is a Gaussian fit to the data.  Visually identical data
  occurs for other redirection probabilities. }
  \label{scaled-width}
\end{center}
\end{figure}

Our simulation results indicate that the random quantity $D_N$ is
self-averaging.  For strictly linear preferential attachment, we find that
the standard deviation grows as $\sqrt{\langle D_N^2 \rangle - \langle D_N
  \rangle^2}\sim N^{1/6}$.  Moreover, the probability distribution $\Pi(D_N)$
of distinct degrees fits the Gaussian
\begin{equation}
\label{Gauss}
\Pi(D_N)=\frac{1}{\sqrt{2\pi \sigma^2}}\,\, e^{-(D_N-\langle D_N\rangle)^2/2\sigma^2}
\end{equation}
extremely well (Fig.~\ref{scaled-width}).  In appropriately scaled
coordinates, this form universally holds for any redirection probability
(equivalently different $\lambda$ values in the attachment rate
$A_k=k+\lambda$).  Moreover, the scaled distributions $f(z)\equiv \sqrt{2\pi
  \sigma^2}\,\Pi(D_N)$, with $z=\sqrt{(\langle D_N^2\rangle -\langle
  D_N\rangle^2)/2\sigma^2}$, are virtually identical for different $\lambda$
values.

\section{The First Hole}

We now study properties of the degree distribution in the sparse regime,
where not every degree is represented.  First consider the location of the
first ``hole'' in the degree distribution---the smallest degree value for
which $N_k=0$.  We define $h_1$ as the degree value of the first hole, $h_2$
as the degree of the second hole, etc.

To determine the location of the first hole, it is useful to use the
probability $P(h)$ that there are no holes in the degree distribution within
the range $[1,h]$.  This coincides with the probability that there is at
least one node of degree $k$ for every $k$ between 1 and $h$.  Again under
the assumption that the number of nodes of degree $k$ is given by an
independent Poisson distribution for each $k$, this probability is given by
\begin{equation}
\label{Poisson:prod}
P(h)  = \prod_{1\leq k\leq h} \left[1-e^{-N_k}\right] \,.
\end{equation}
We estimate the location of the first hole from the criterion $P(h_1) =
\frac{1}{2}$; however, any constant between 0 and 1 could equally well be
chosen in this condition.  Taking the logarithm of \eqref{Poisson:prod} and
using $\ln\left[1-e^{-N_k}\right]\approx -e^{-N_k}$ (which is justifiable
since $e^{-N_k}\ll 1$ when $k\leq h_1$), gives the following for the average
location $\langle h_1\rangle$ of the first hole:
\begin{equation}
\label{first}
\int_1^{\langle h_1\rangle} dk\,e^{-N_k} = \ln 2\,.
\end{equation}
Using Eq.~\eqref{Nk} in \eqref{first} we find
\begin{subequations}
\begin{equation}
\label{h1}
\langle h_1\rangle \simeq \left(\frac{\nu N R}{\ln H}\right)^{1/\nu}, \qquad 
H \sim \frac{\nu N R}{[\ln(\nu N R)]^{1+\nu}}~.
\end{equation}
Since $H$ appears inside the logarithm, one can ignore the logarithmic factor
in $H$ itself, thereby giving the simpler and still asymptotically exact
formula
\begin{equation}
\label{h1:simple}
\langle h_1\rangle \simeq \left(\frac{\nu N R}{\ln N}\right)^{1/\nu}~.
\end{equation}
\end{subequations}

\begin{figure}[ht]
\begin{center}
\includegraphics[width=0.5\textwidth]{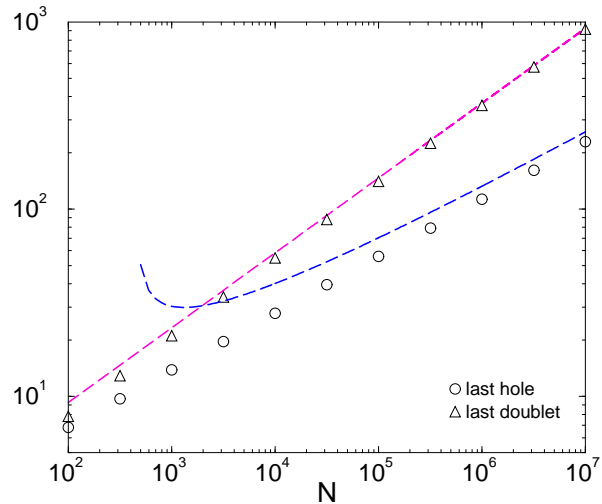}
\caption{Location of the last hole and the last doublet as a function of $N$
  for $10^4$ realizations of networks that are grown by strictly linear
  preferential attachment.  The dashed curve is the prediction \eqref{h1},
  while the straight dashed line is the prediction from Eq.~\eqref{V}.}
  \label{hole-doublet}
\end{center}
\end{figure}

It is worth noting that the naive calculation that leads to \eqref{D_naive}
for the number of distinct degrees ignores the possibility that holes exist
in the range $k\leq (NR)^{1/\nu}$.  According to Eq.~\eqref{h1:simple},
however, the first hole appears earlier than $(NR)^{1/\nu}$ in the
$N\to\infty$ limit.  For a terrestrial-scale network with, say $N=10^9$
nodes, the location of the first hole will be roughly 3 times smaller than
that predicted by the naive estimate \eqref{D_naive}.

\section{Last Doublet and Last Dimer}

Somewhere in the tail of the degree distribution lies the last doublet, the
largest two consecutive $k$ values for which $N_k>0$, and the last dimer, the
largest $k$ value for which $N_k=2$ (Fig.~\ref{sketch-10000000}).  Starting
with degree 1, the degree distribution first consists of a long string of
consecutive ``occupied'' degrees $1\leq k<h_1$, followed by a second string
in the degree range $h_1< k<h_2$, etc.  As the degree increases, these
strings become progressively shorter and above a certain threshold all
remaining strings are singlets.  For a large network, the last string that is
not a singlet will almost certainly be a doublet (with probability
approaching 1 as $N\to\infty$).  We now determine the average position of
this last doublet.

The probability to have a doublet at $(k,k+1)$ is $N_k^2$ when $k\gg K=
(NR)^{1/\nu}$.  To estimate the position of the last doublet
$(\delta,\delta+1)$ we employ the extremal criterion
\begin{equation}
\label{criterion}
\sum_{k\geq \delta}N_k^2 \sim 1
\end{equation}
that there should be of the order of one doublet in the degree range
$(\delta,\infty)$.  Using $N_k\sim NR\,k^{-\nu}$, we obtain
\begin{subequations}
\label{deltas}
\begin{equation}
\label{delta_av}
\langle \delta\rangle = C (RN)^{1/(\nu -1/2)}\,,
\end{equation}
with $C$ a constant, for the average position of the last doublet.  Notice
that the position of the last doublet also coincides, up to a prefactor of
the order of 1, to the position of the last dimer.  A more precise approach
to determine the average location of the last doublet gives the amplitude as
\begin{equation}
\label{delta_C}
C = (2\nu-1)^{ - 1/(2\nu-1)}\Gamma\!\left(\frac{2\nu-2}{2\nu-1}\right)~.
\end{equation}
\end{subequations}
To establish \eqref{delta_C}, we use the the independent Poisson
approximation to write, for the probability $F(\delta)$ to have no doublets
in the degree range $k>\delta$,
\begin{equation}
\label{F}
F(\delta)=\prod_{k> \delta}\left[1-\left(1-e^{-N_k}\right)^2\right]\,.
\end{equation}
This expression is the straightforward generalization of
Eq.~\eqref{Poisson:prod} to the case of dimers.  Since the average number of
nodes with degrees in the range $k>\delta$ is small, the product on the
right-hand side of \eqref{F} simplifies to $\exp\!\left[ -\int_\delta^\infty
  dk\,N_k^2\right]$. Computing the integral gives
\begin{equation}
\label{Fk}
F = \exp[-(\delta_0/\delta)^{2\nu-1}], \qquad\qquad \delta_0 =  \left(\frac{R^2N^2}{2\nu\!-\!1}\right)^{1/(2\nu\!-\!1)}\hspace{-5ex}.
%F(\delta) = \exp\!\left[ -\left(\frac{\delta_0}{\delta}\right)^{2\nu-1}\right]
\end{equation}
The probability density $\Phi=\frac{dF}{d\delta}$ for the last doublet is
then
\begin{equation}
\label{Phi}
\Phi(\delta) = \frac{2\nu-1}{\delta}\left(\frac{\delta_0}{\delta}\right)^{2\nu-1}
\exp\!\left[ -\left({\delta_0}{\delta}\right)^{2\nu-1}\right]\,,
\end{equation}
from which the average position of the last doublet is given by
\begin{subequations}
\begin{equation}
\label{delta_comp}
\langle \delta\rangle = \int_0^\infty d\delta\,\,\delta\,\Phi(\delta)
= \int_0^\infty d\delta\,[1-F(\delta)]\,.
\end{equation}
Substituting \eqref{Fk} into \eqref{delta_comp} leads to
\begin{equation}
\langle \delta\rangle = \Gamma\!\left(\frac{2\nu-2}{2\nu-1}\right)  \delta_0\,.
\end{equation}
\end{subequations}
which reproduces \eqref{deltas}. Similarly, the mean-square position of the
  last doublet is
\begin{equation}
\label{delsq}
\langle \delta^2 \rangle = \int_0^\infty d\delta\,\,\delta^2\,\Phi(\delta)
= 2\int_0^\infty d\delta\,\delta\,[1-F(\delta)]\,,
\end{equation}
from which the variance is
\begin{equation}
\label{delvar}
\langle \delta^2 \rangle - \langle \delta\rangle^2 = 
\left[\Gamma\!\left(\frac{2\nu-3}{2\nu-1}\right) - \Gamma^2\!\left(\frac{2\nu-2}{2\nu-1}\right)\right] \delta_0^2\,.
\end{equation}
For strictly linear preferential attachment network growth, the above results
reduce to
\begin{align}
\begin{split}
\label{V}
&\langle \delta\rangle = AN^{2/5}, \qquad ~~\,\, A=(\tfrac{16}{5})^{1/5}\Gamma(\tfrac{4}{5})= 1.469158\ldots \\
&{\langle \delta^2 \rangle^{1/2}} = V\langle \delta\rangle, \qquad
\,V=\frac{\sqrt{\Gamma(3/5)}}{\Gamma(4/5)} =                    1.048182\ldots
\end{split}
\end{align}

Following the same line of reasoning, the position of the last triplet,
$(\tau-1, \tau, \tau+1)$, is given by
\begin{equation}
\label{triplet_av}
\langle \tau\rangle \sim N^{1/(\nu -1/3)}\,.
\end{equation}
For strictly linear preferential attachment, this result gives the dependence
$\langle \tau\rangle \sim N^{3/8}$.  Our simulation data are consistent with
the predictions \eqref{V}--\eqref{triplet_av} (Fig.~\ref{hole-doublet}).

\section{Discussion}

For any broadly distributed integer-valued variable, the underlying
distribution exhibits intriguing features that stem from the combined
influences of discreteness and finiteness.  Such a distribution is smooth in
a dense regime, where every integer value of the variable has a non-zero
probability of occurrence.  In the complementary sparse regime, a variety of
statistical anomalies arise that quantify the extent of the
sparseness~(Fig.~\ref{sketch-10000000}).

For the degree distribution of complex networks that genererically have
power-law tails, $N_k\sim N/k^\nu$, our main results are: (i) The number of
distinct degrees in a network of $N$ nodes scales as $N^{1/\nu}$.  This
generic behavior is also observed in the citation network of the Physical
Review.  (ii) The distribution in the number of distinct degrees is very well
fit by a universal Gaussian function.  (iii) There is a rich set of behaviors
for basic characteristics of the sparse regime, such as the positions of
holes (zeros) in the distribution, as well as the locations of doublet,
triplets, etc., and the locations of dimers, trimers, etc.  All of these
quantities can be determined by simple probabilistic reasoning.

Our analysis tacitly assumed that the number of nodes of different degrees,
$N_i$ and $N_j$ for $i\ne j$, are uncorrelated, and that the $N_k$'s are
Poisson distributed random quantities.  While these assumptions are
questionable in the sparse regime, predictions that are based on these
assumptions are in excellent agreement with results from simulations of
preferential attachment networks.  While we believe that our predictions are
asymptotically exact, a more rigorous analysis is needed to justify them and
explain their validity (or at least their impressive accuracy).  A
challenging extension of this work is to probe the fluctuations in the total
number of distinct degrees.  The mechanism for the observed Gaussian shape of
the distribution of distinct degrees is not at all evident.  In fact, for
networks that grow by redirection, with redirection becoming more certain as
the degree of the ancestor node increases, the total number of distinct
degrees is not even a self-averaging quantity~\cite{GKR_in}.

Our methods apply equally well to other heavy-tailed integer-valued
distributions, such as the cluster-size distribution in classical
percolation~\cite{AS91} and in protein interaction and regulatory
networks~\cite{expt}.  The latter models often exhibit an infinite-order
percolation transition, in which the cluster-size distribution has an
algebraic tail in the entire non-percolating phase
\cite{clusters,sam,kr,3kr,DL02,bb,cb,kb}.  Our approach leads to new results
for the total number of distinct cluster types $C_N$, for the position of the
first hole (the minimal size that is not present), etc.

For concreteness, consider networks that are built by adding nodes one at a
time with each new node connecting to $k$ randomly chosen existing nodes with
probability $p_k$~\cite{cb,kb}.  While the set of probabilities $p_k, k=0, 1,
2,\ldots$, with $\sum_{k\geq 0}p_k=1$, fully defines the network ensemble,
only the first two moments, $\langle k\rangle = \sum_{k\geq 1} kp_k$ and
$\Delta= \langle k^2\rangle - \langle k\rangle ^2$, matter in determining
large-scale properties.  In the non-percolating phase, $\langle k\rangle <
\frac{1}{2}$ and $\Delta < \frac{1}{4}$, we use the decay exponent for the
cluster-size distribution that was determined in \cite{kb} to obtain
\begin{equation*}
C_N \sim N^\delta, \qquad\qquad \delta = \frac{1-\sqrt{1-4\Delta}}{3-\sqrt{1-4\Delta}}~.
\end{equation*}

At the percolation transition, $\langle k\rangle < \frac{1}{2}$ and $\Delta =
\frac{1}{4}$, the tail of the cluster-size distribution contains universal
(independent of $\langle k\rangle$ and $\Delta$) algebraic and logarithmic
factors, viz.\ $c_s\simeq 2(1-2\langle k\rangle)^{-2}s^{-3}(\ln s)^{-2}$ for
$s\gg 1$.  A straightforward generalization of our previous analysis shows
that the total number of distinct cluster types grows as
\begin{equation*}
  C_N \simeq 2^{1/3}\, \Gamma\big(\tfrac{2}{3}\big)\,  
  \Big(\frac{3}{1-2\langle k\rangle}\Big)^{2/3}\left[\frac{N}{(\ln N)^2}\right]^{1/3}.
\end{equation*}

As a final note, this work has focused broadly on properties associated with
the support of discrete distribution.  The averages of these properties over
a large ensemble of networks have systematic dependences on the number of
nodes $N$ in the network; however, the behavior in each network realization
may not be monotonic.  Thus while $k_{\rm max}$ is clearly a non-decreasing
function of $N$, the number of distinct degrees and the locations of
quantities like the first hole or the last doublet can both increase or
decrease with $N$.  This intriguing aspect of the problem may provide a more
detailed understanding of how a complex network actually grows.

\bigskip\noindent We gratefully acknowledge partial financial support from
AFOSR and DARPA grant \#FA9550-12-1-0391, and from NSF grant No.\
DMR-1205797.  We also thank A. Gabel for collaborations at a preliminary
stage of this project.  Finally, we thank the American Physical Society
editorial office for providing the citation data.  
\end{document}